\documentclass[aps,twocolumn,showpacs,groupedaddress,floatfix]{revtex4-1}
\usepackage{graphicx}
\usepackage{color}

\newcommand{\micron}{\ensuremath{\mu{\mbox{m}}}}
\newcommand{\Vd}{\ensuremath{V_\mathrm{D}}}
\newcommand{\nosp}{\!\!}

\begin{document}

\title{Dissipative Transport of a Bose-Einstein Condensate}

\author{D. Dries, S. E. Pollack, J. M. Hitchcock, and R. G. Hulet}
\affiliation{Department of Physics and Astronomy and Rice Quantum
Institute, Rice University, Houston, Texas 77005, USA}

\date{\today}

\begin{abstract}

We investigate the effects of impurities, either correlated disorder or a
single Gaussian defect, on the collective dipole motion of a
Bose-Einstein condensate of $^7$Li in an optical trap. We find that
this motion is damped at a rate dependent on the impurity
strength, condensate center-of-mass velocity, and interatomic
interactions. Damping in the Thomas-Fermi regime depends universally on the
disordered potential strength scaled to the condensate chemical
potential and the condensate velocity scaled to the speed of
sound.  The damping rate is comparatively small in the weakly interacting regime, and 
in this case, is accompanied by strong condensate fragmentation.
\textit{In situ} and time-of-flight images of the atomic cloud
provide evidence that this fragmentation is driven by dark soliton
formation.

\end{abstract}


\maketitle

\section{Introduction\label{sec:Intro}}

The creation of Bose-Einstein condensates (BECs) of ultracold atomic gases \cite{Anderson95,Bradley95,Davis95}
has enabled investigations of some of the most fundamental concepts of condensed matter physics \cite{bloch08}.
One of the most fruitful avenues of research has involved the use of BECs to probe the nature of superfluidity
itself. Early studies led to observations of the critical velocity for the onset of dissipation
\cite{Raman99,onofrio00,Raman01} and quantized vortices \cite{Madison2000,Abo-Shaeer01,haljan01,hodby01}.

Recently, there has been much interest in using BECs to emulate disordered
superfluids (c.f. \cite{fallani08,SanchezP2010}). Results from such experiments have wide ranging implications, from the transport of superfluid He in
porous media \cite{reppy92} to the motion of atomic BECs in microchip
traps or matter waveguides \cite{Fort02,Leanhardt03,Jones03,Aspect04}. Of particular
interest is how disorder can disrupt, or even destroy,
superfluidity. Due to their exquisite controllability, atomic BECs
are ideal physical systems with which to systematically study the
interplay between superfluidity, disorder, and interatomic interactions.

In this paper, we report measurements of the dissipation of the superfluid flow of an elongated BEC subject to
either a disordered potential or a single Gaussian defect. We characterize the superfluid nature of the
harmonically trapped cloud through detailed measurements of the velocity dependent damping of the collective
dipole mode. We use a BEC of $^7$Li in the $|F=1, m_F=1\rangle$ internal state, where the interactions may be
tuned via a wide Feshbach resonance located at 737\,G \cite{streckerSolitons, junker08, pollack09}. This
resonance includes a shallow zero-crossing that enables the $s$-wave scattering length $a$ to be tuned over a
range of nearly 7 decades, with $a$ as small as $0.01\,a_0$, where $a_0$ is the Bohr radius \cite{pollack09}. The
gas may be made nearly ideal with transport properties strikingly different from the more strongly interacting
case. Furthermore, the healing length $\xi=1/\sqrt{8\pi n_0 a}$, where $n_0$ is the peak density of the
condensate, may be made as large as the condensate itself. In this regime, effects due to the fundamental wave
nature of individual atoms become important. For example, if $\xi$ is on the order of the disorder grain size or
larger, a BEC can become an Anderson localized insulator \cite{aspectAL,modugnoAL}. In addition, the chemical
potential $\mu$ in this weakly interacting regime may be less than the radial harmonic oscillator ground state
energy, effectively ``freezing out'' the radial dynamics and leading to quasi-one-dimensional behavior.

\subsection{Superfluidity of a BEC}\label{sec:becsuper}

One of the seminal results originating from the theory
of superfluid $^4$He is Landau's criterion.  According to this criterion, elementary excitations can be created only if the fluid velocity $v$
is greater than Landau's critical velocity $v_L$ \cite{landau41,tilleybook}
\begin{equation}
v_L=\min \frac{\epsilon(p)}{p},
\label{eqn:landau}
\end{equation}
where $\epsilon(p)$ is the energy of an elementary excitation of momentum $p$.
For the case of a weakly interacting BEC with uniform density $n$,
Bogoliubov theory gives the excitation energy as \cite{pethickbook}
\begin{equation}
\epsilon (p) =\sqrt{ \left(\frac{p^2}{2 m}\right)^2 +  c^2p^2},
\label{eqn:spec}
\end{equation}
where $m$ is the atomic mass and $c$ is the bulk speed of sound. For small $p$,
this spectrum reduces to the well known relation $\epsilon (p) = c\,p$ describing phonon excitations with
\begin{equation}
c=\sqrt{\frac{n U}{m}}, \label{eqn:usound}
\end{equation}
where $U=4\pi\hbar^2a/m$.  Application of Eq.~\ref{eqn:landau} 
gives $v_L=c$, implying that only supersonic flow
can dissipate energy through the creation of elementary excitations;
conversely, if the flow is subsonic, excitations are energetically
forbidden, and the flow is superfluid. Application of
Eq.~\ref{eqn:landau} to the case of a non-interacting condensate
implies that $v_L=0$, suggesting that superfluidity cannot exist in
an ideal gas.

The dynamics of highly elongated BECs can be
accurately modeled using an effective one-dimensional (1D) nonlinear Schr\"{o}dinger
equation (NLSE)
\cite{zaremba98,Kavoulakis98,stringari98,salasnich02}. In such a
treatment, one starts from the 3D Gross-Pitaevskii equation (GPE),
and integrates out the radial dimension.  The effect of this
integration is a reduction in $c$ relative to Eq.~\ref{eqn:usound}
due to the average over the nonuniform radial density.  For
the case of a harmonically trapped BEC in the Thomas-Fermi regime,
the bulk density $n$ is replaced with the average density $n_0/2$. 
Therefore, the speed of sound becomes
\begin{equation}
c_0 = \sqrt{ \frac{n_0 U}{2 m} }. \label{eqn:sound}
\end{equation}
A theoretical description of an elongated BEC beyond the standard 1D NLSE leads to a reduction (on the order of
$10\%$) to the speed of sound relative to Eq.~\ref{eqn:sound}~\cite{kamchatnov04}.  In addition, the spectrum of
axially propagating excitations in a cylindrical BEC can differ dramatically from Eq.~\ref{eqn:spec} when
$\mu\gg\hbar\omega_r$, where $\mu$ is the chemical potential, leading to an additional reduction in $v_L$~\cite{fedichev01}.  The highest $\mu$
condensates created in our system have $\mu/\hbar\omega_r\sim13$, resulting in a predicted $20\%$ reduction~\cite{fedichev01}.

When attempting to explain the onset of dissipation in any particular experimental situation, care must be taken
to apply Landau's criterion \textit{locally}, by using the local density $n(r=0,z)$ instead of $n_0\equiv
n(r=0,z=0)$ in Eq.~\ref{eqn:sound} \cite{frisch92}. For arbitrary trapping potentials, excitations will be
nucleated first in regions of low density where the local speed of sound is small, and the critical velocity is
reduced relative to the bulk.  As a consequence of this effect, experimentally observed critical velocities are
often much lower than the bulk speed of sound \cite{Raman99,onofrio00,Raman01}.

\bigskip
The remainder of this paper is organized as follows: In
Sec.~\ref{sec:Exp} we describe our experimental methods for creating
a BEC in either a disordered harmonic potential or a harmonic
potential with a single Gaussian defect; in Secs.~\ref{sec_dis} and
\ref{sec_bar} we discuss our results for the induced dissipation for
these two scenarios, where both
the 3D Thomas-Fermi and the quasi-1D weakly interacting regimes
are discussed for each case.
We conclude in Sec.~\ref{sec_conc} with a discussion relating the
similarities and differences between dissipation in the two types of
potentials, and directions for future studies.

\section{Experimental Method}\label{sec:Exp}

We create a BEC of $^7$Li in a highly elongated, cylindrically symmetric, 
hybrid magnetic-optical dipole trap~\cite{chen08, pollack09} with radial and axial trapping frequencies in the ranges of 
$\omega_r/\left(2 \pi\right)\sim \,220$--$460$\,Hz and $\omega_z/\left(2 \pi\right)\sim\,4$--$5.5$\,Hz, respectively. The radial
confinement is dominated by the optical trapping potential formed by a single focused laser beam with wavelength
1030\,nm and a $1/e^2$ Gaussian radius of $33\,\micron$, while the axial confinement is dominated by an
adjustable, harmonically confining magnetic field.  A set of Helmholtz coils provides a uniform bias field along
the long ($z$) axis of the trap, allowing for the tuning of $a$ via a Feshbach resonance at 737\,G
\cite{streckerSolitons,junker08,pollack09}. The BEC is created at a field of 717\,G where $a$ is positive and
large enough ($\sim\nosp200\,a_0$) to allow for efficient evaporative cooling in the optical trap, but small
enough to avoid substantial three-body losses.  At this field, the trap lifetime is limited to $\sim\nosp10\,$s
due to three-body recombination with a loss coefficient of $L_3\sim10^{-26}$\,cm$^6$/s \cite{dries09}. After
evaporation, the BEC has no discernible thermal component from which we estimate that the temperature $T <
0.5\,T_C$, where $T_C$ is the BEC transition temperature. The bias field is then ramped over a timescale on the
order of seconds to achieve the desired value for $a$.

We excite the collective dipole mode of the condensate by pulsing on an
axially oriented magnetic gradient, thereby abruptly shifting the
center of the harmonic trap.  After $1/4$ of an
oscillation period, the condensate is at the peak of an
oscillation and we abruptly switch on either a disordered
potential with an extent exceeding the
oscillation amplitude of the condensate, or a single, narrow Gaussian
defect located near the trap center. By varying the duration of the
gradient pulse we precisely vary the amplitude $A$ of the
oscillation, and therefore the initial peak velocity $v_0$ of the
condensate center of mass, where $v_0=A\,\omega_z$. At various times
thereafter we image the cloud to track the center of mass location
as well as the shape of the density distribution.
We investigate the dependence of the damped dipole motion on $v_0$, the
strength of the disordered potential or single Gaussian defect, and on
the value of $a$.

\begin{figure}
\includegraphics[angle=-90,width=1.0\columnwidth]{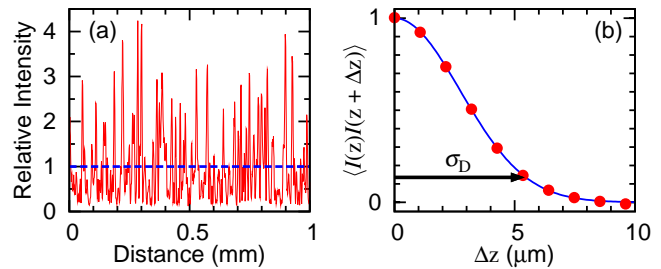}
\caption{(Color online) Disordered potential created from laser speckle. (a) Cut
through an image of the speckle potential. The disorder strength
$\Vd$ is proportional to the average value of the intensity $\left<I\right>$ (dashed
line). (b) The autocorrelation of the intensity distribution is well
fit by a Gaussian with $1/e^2$ radius $\sigma_\mathrm{D} =
5.5\,\micron$.  For some of the data in this paper
(Figs.~\ref{fig:powerlaw},~\ref{fig:damping}, and
~\ref{fig:dampva.ps}) $\sigma_\mathrm{D}=3.4\,\micron$. }
\label{fig:disfig}
\end{figure}

The disordered potential is an optical speckle pattern created by passing a laser beam through a diffuser plate
in a manner similar to previous studies \cite{chen08, lye05, clement05,schulte05}. This beam is directed
perpendicular to the trap $z$-axis. Figure~\ref{fig:disfig} shows a characteristic intensity slice of the
disorder. The disorder speckle size $\sigma_\mathrm{D}$ is defined to be the $1/e^2$ radius of a Gaussian fit to
the autocorrelation of the intensity pattern and is measured to be $\sigma_\mathrm{D}=5.5\,\micron$. 
The beam has been cylindrically focused such that in the
radial direction the speckle size is much larger than the radial Thomas-Fermi radius
$\sim\nosp10\,\micron$, making the disorder effectively 1D.  We have verified that the
intensity distribution of the disorder follows a decaying exponential $P(I) = \left<I\right>^{-1}
e^{-I/\left<I\right>}$, as expected for fully developed speckle \cite{goodman}.  The average value of the speckle
intensity $\left<I\right>$ determines the disorder strength through the relation $\Vd=\hbar\,\Gamma^2
\left<I\right>/(4I_\mathrm{sat} \Delta)$, where the transition linewidth $\Gamma=(2\pi)\,5.9\,\mathrm{MHz}$ and the
saturation intensity $I_\mathrm{sat}= 5.1\,\mathrm{mW/cm}^2$.  The detuning from the $^7$Li
$2S\rightarrow{2P}$ transition $\Delta=(2\pi)\,300\,\mathrm{GHz}$, producing a repulsive disorder
potential. For the strongest disorder used in these studies, off-resonant scattering from the disorder occurs at
a rate of $\sim\!0.1$\,s$^{-1}$. The statistical properties of the speckle pattern are measured by direct imaging
with a CCD camera before the optical system is installed onto the experimental apparatus.

A cylindrically focused laser beam is used for the studies involving a single Gaussian defect.
This beam has a Gaussian
intensity distribution
$I\left(z,r\right)=I_0e^{-2(r^2/w_r^2 + z^2/w_z^2)}$, with beam waists
$w_r=5\,$mm and $w_z=12\,\micron$.  The radial size of the defect $w_r$ is much larger than $R_\mathrm{TF}$,
ensuring that flow around the defect is suppressed.
We conduct experiments using both a repulsive (blue detuned)
and an attractive (red detuned) defect with $|\Delta|=300\,$GHz.

We adjust the healing length through an approximate range
$0.5\,\mu\mbox{m} < \xi < 20\,\mu\mbox{m}$ by tuning $a$. Thus, a
wide range of values are achievable for the relevant dimensionless
quantities, $0.1< \xi/\sigma_\mathrm{D} < 3.6$ and $0.04 < \xi/w_z <
1.7$.

\section{Disorder Induced Dissipation}\label{sec_dis}
\subsection{Thomas-Fermi Regime}
\begin{figure}
\includegraphics[angle=-90,width=1.0\columnwidth]{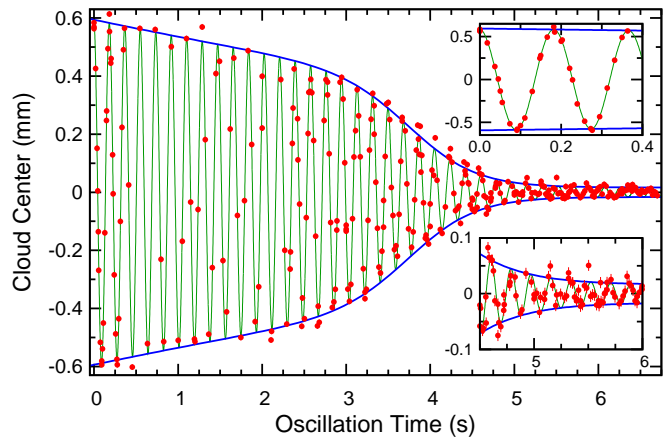}
\caption{(Color online) Damping of a condensate initially traveling supersonically through a disordered potential with $\Vd/h =
280\,\mathrm{Hz}$. The center of the BEC (circles) is extracted from a Thomas-Fermi fit to the radially
integrated column density (the ``axial density''). The thick lines tracing the amplitude are phenomenological
guides to the eye. The initial amplitude is $A = 0.6\,$mm yielding an initial peak velocity of $v_0 =
20\,\mathrm{mm/s}$. For this data, $\omega_z=(2\pi)\,5.5\,$Hz, $\omega_r = (2\pi)\,260\,$Hz, $a = 25\,a_0$, and
$\mu = \frac{1}{2}m\omega_z^2 R_\mathrm{TF}^2 = h\,(1.1\,\mathrm{kHz})$, where $R_\mathrm{TF}$ is the axial Thomas-Fermi radius. In addition, $c_0 = 5.6\,\mathrm{mm/s}$,
$\xi = 0.8\,\micron$, and $\xi/\sigma_\mathrm{D} = 0.2$. The insets show details of the oscillation at early and
late times. } \label{fig:fancyDecay}
\end{figure}
\begin{figure}
\includegraphics[angle=-90,width=1.0\columnwidth]{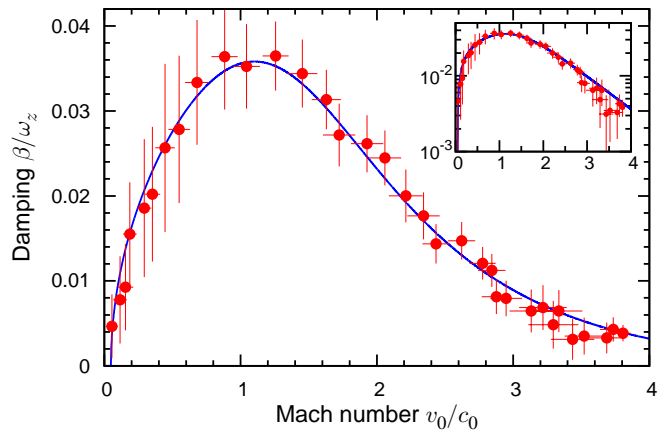}
\caption{(Color online) Velocity dependent damping. Results of fitting the data of
Fig.~\ref{fig:fancyDecay} to Eq.~\ref{eqn:osc} using a traveling
4-period window.  The peak velocity $v_0$ is obtained from
$v_0=A\,\omega_z$. The solid line is a square-root function convolved
with an exponential decay and is meant as a guide to the eye. The
inset shows the same data on a semi-log plot, emphasizing the nearly
exponential decay of $\beta/\omega_z$ for large $v_0/c_0$. Vertical
error bars correspond to the range in $\beta$ for which
$\Delta\chi^2=1$ for the fit to Eq.~\ref{eqn:osc} while
simultaneously adjusting $A$, $\beta$, and $\phi$ to minimize $\chi^2$.
Horizontal error bars are determined using an identical process
for $A$ in Eq.~\ref{eqn:osc} and are typically $\sim\nosp15\%$.
The effects of systematic uncertainty in imaging magnification and
variations in $N$ are $\sim\nosp10\%$ in the horizontal axis
and $\sim\nosp5\%$ in the vertical axis, these
are not included in the displayed error bars.
} \label{fig:vdamp}
\end{figure}

Figure~\ref{fig:fancyDecay} shows the position of the center of a
condensate at various times during a dipole oscillation in a
disordered potential. The dipole oscillation is initiated by a kick
that produces an initial peak velocity of $v_0= 20$\,mm/s when the
condensate passes through the center of the trap. For this data, the
condensate begins its motion well into the supersonic regime with
$v_0 \sim 4\,c_0$. The resulting oscillation is characterized by a
time-dependent damping, suggesting that the damping depends on
$v_0$. The damping rate is initially small, goes through a
maximum after about $3.8\,$s, and then diminishes at later times.
We fit 4-period sections of the data in Fig.~\ref{fig:fancyDecay} to
the form of a damped harmonic oscillator:
\begin{equation}\label{eqn:osc}
z(t)=A e^{-\beta t} \cos\left(\omega' t+\phi\right),
\end{equation}
where $\omega'=(\omega_z^2-\beta^2)^{1/2}$.  The peak velocity $v_0$ is then computed from the fitted $A$ for
each data subset to obtain the damping coefficient $\beta$ as a function of $v_0$, with the results shown in
Fig.~\ref{fig:vdamp}. The damping monotonically increases for small $v_0$, peaking near $v_0 \sim 1.1\,c_0$,
followed by a nearly exponential decay of $\beta$ for $v_0 > c_0$.

A perturbative theoretical treatment has produced a closed
form solution for the velocity-dependent damping, resulting in good
quantitative agreement with our measurements \cite{bhongale10}. For
weak disorder the qualitative behavior shown in Fig.~\ref{fig:vdamp}
can be understood through a local Landau critical velocity argument.
At low velocities, Bogoliubov quasiparticles
are only created within a thin shell at the surface of the
condensate, where the low density leads to a low local speed of
sound, and therefore a low local $v_L$.  As the
velocity of the condensate increases, a larger condensate volume
can support excitations because a larger fraction of the atoms
violate the local Landau criterion. The maximum damping occurs near
the point where the velocity of the BEC reaches the peak speed of
sound $c_0$ in the condensate. At even larger velocities the
excitation volume cannot increase further, but the Bogoliubov
density of states decreases, resulting in the observed exponential
decrease of the damping.

Except for the absence of a critical velocity, the qualitative behavior of the velocity dependent damping shown
in Fig.~\ref{fig:vdamp} is remarkably similar to that predicted by 1D NLSE simulations of a uniform, repulsive
BEC in the presence of an oscillating Gaussian obstacle \cite{Radouani03,Radouani04}.
In these simulations, above a certain impurity strength-dependent critical velocity, the impurity moving at a
velocity $v$ deposits energy into the BEC in the form of density fluctuations. The average rate of condensate
energy growth $\left<dE/dt\right>$ increases nearly linearly with $v$, to a peak at $v \sim c$ as the defect
excites dark solitons and linear sound waves. As the velocity of the defect is increased further, the density
fluctuations decrease significantly, accompanied by an exponential decrease of $\left<dE/dt\right>$, similar to
our experimental observations.  
In contrast to a single impurity in a uniform condensate,
a defect is always present in a low density region of a condensate in a disordered harmonic trap.
Consequently, $v_0$ is always greater than the local speed of sound at the edge of the
condensate and excitations are always present. Previous experimental \cite{lye05,modugno06,chen08} and numerical
\cite{albert08} studies of the damping of collective modes and the damping of Bloch oscillations in a disordered
lattice potential \cite{schulte08,drenkelforth08} have found qualitatively similar results.

\begin{figure}
\includegraphics[width=1.0\columnwidth]{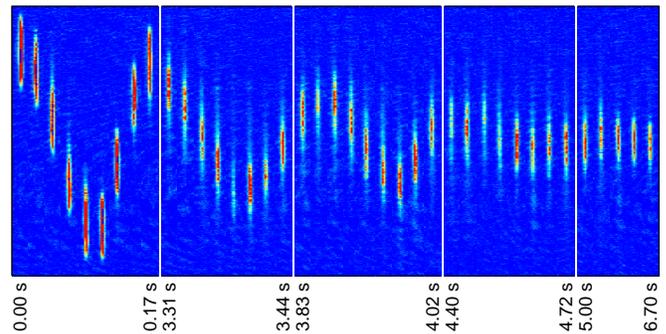}
\caption{(Color online) Characteristic \textit{in situ} polarization phase-contrast images of the data shown in
Fig.~\ref{fig:fancyDecay} at various times. The images are nearly equally spaced in time between the time labels.
}\label{fig:2dfancy}
\end{figure}
\begin{figure}
\includegraphics[angle=-90,width=1.0\columnwidth]{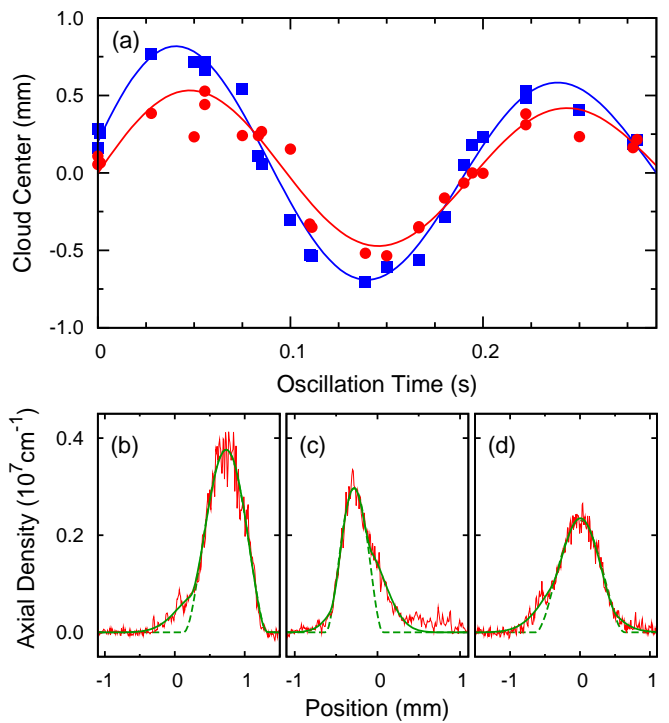}
\caption{(Color online) Generation of a non-condensed component. (a)~Squares show the center of the Thomas-Fermi (condensed)
component and circles show the center of the Gaussian (non-condensed) component. The Gaussian center trails
behind the Thomas-Fermi center and has a smaller amplitude of oscillation. Within experimental uncertainty,
$\omega_z=(2\pi)\,5.1(2)$\,Hz for both components. For this data, $a = 200\,a_0$, $N=3\times10^5$,
$\mu/h=1.8$\,kHz, $\Vd/\mu=0.22$, $v_0 = 28\,$mm/s, $c_0 = 7.2\,$mm/s, and $\omega_r=(2\pi)\,220$\,Hz. (b--d)
Axial density distributions with bimodal fits (solid lines) and a single component Thomas-Fermi fit (dashed
lines) at various times during the oscillation: (b) 28\,ms, (c) 100\,ms, (d) 190\,ms. The condensates in (b) and
(d) are traveling in the positive direction whereas the condensate in (c) is traveling in the negative
direction.}\label{fig:bimodal}
\end{figure}

Figure~\ref{fig:2dfancy} shows \textit{in situ} polarization phase-contrast images \cite{bradley97} of the BEC at
various times in the oscillation shown in Fig.~\ref{fig:fancyDecay}. The damping clearly does not result from a
loss of collectivity as predicted by 1D NLSE numerical simulations \cite{albert08}. Rather, the BEC nearly
maintains its original shape throughout the oscillation. Close inspection of the density distributions in
Fig.~\ref{fig:2dfancy} reveals a ``tail'' of non-condensed atoms that appears to oscillate slightly out-of-phase
with the central Thomas-Fermi distribution.  At early times, these non-condensed atoms appear to lag behind the
BEC, while at later times they oscillate in-phase with it. This two-component out-of-phase oscillation is
reminiscent of the second sound-like oscillation reported in Ref.~\cite{meppelink09}. In that work, the initial
temperature was high enough that damping occurred due to the interaction between a BEC and a thermal component.
In contrast to those results, we observe that the dipole oscillation is undamped in the absence of the disordered
potential. Furthermore, there is no observable heating due to the quick switch on of the disorder.
In our experiment, therefore, the presence of the non-condensed component seems
to be linked to the motion of the BEC in the disordered potential.  A recent numerical simulation using a truncated Wigner method predicts the emission of incoherent atoms from a
BEC moving supersonically through a disordered potential \cite{scott08}, consistent with our observations.

We have investigated this effect in further detail using \textit{in situ} absorption imaging, which allows for
determination of the low density non-condensed wings of the distributions. Figure~\ref{fig:bimodal} shows that by
fitting the cloud to a bimodal Thomas-Fermi plus Gaussian profile, a phase difference of $\Delta\phi=0.23$
between the condensed and non-condensed cloud centers is found.  Note that the interaction strength is
different for this data than that shown in Figs.~\ref{fig:fancyDecay}--\ref{fig:2dfancy}.

\begin{figure}
\includegraphics[angle=-90,width=1\columnwidth]{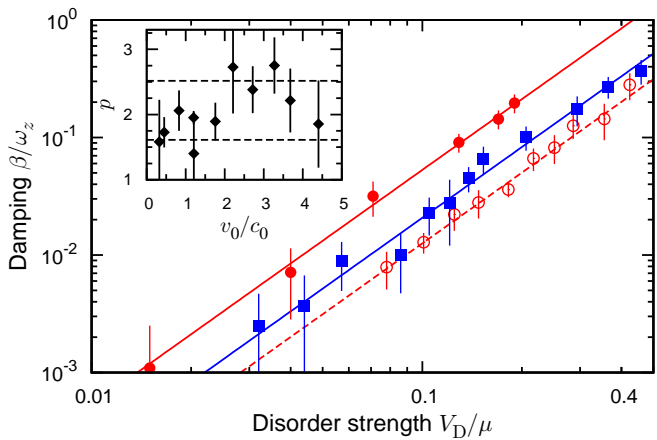}
\caption{(Color online) Damping vs.~$\Vd$. Open circles correspond to the data shown in Fig.~\ref{fig:phasedis} ($a = 200\,a_0$)
in the range $0.7<v_0/c_0<0.9$; filled circles correspond to $a = 200\,a_0$, $v_0/c_0 = 1.2$, $\mu / h =
2.2\,$kHz; squares correspond to $a = 25\,a_0$, $v_0/c_0 = 1.2$, $\mu / h = 750\,$Hz.  The damping parameter
$\beta$ follows a power law with $p\sim2$ (solid and dashed lines), independent of $\mu$ or $a$. To minimize
systematic effects associated with the velocity dependence of $\beta$ (e.g., Figs.~\ref{fig:fancyDecay} and
\ref{fig:vdamp}), we fit a 4-period window for which the data is described well by Eq.~\ref{eqn:osc}. Vertical
error bars are as defined in Fig.~\ref{fig:vdamp}.  The inset shows the fit values of $p$ as a function of
$v_0/c_0$ for a collection of data sets at $a = 200\,a_0$. The dashed lines indicate the plus-and-minus one
standard deviation extent for the collection of measured velocities. Vertical error bars for $p$ are determined
as in Fig.~\ref{fig:vdamp} using a fit to Eq.~\ref{eqn:powerlaw} for each oscillation at a given $v_0/c_0$.  Data
corresponding to filled circles and squares was taken using an optical trap setup different from that described
in Sec.~\ref{sec:Exp} with $\lambda=1064$\,nm and a beam waist of $24\,\micron$ resulting in
$\omega_z=(2\pi)\,4.9$\,Hz, $\omega_r=(2\pi)\,460$\,Hz, and $N=3\times10^5$. Also, for these data sets
$\sigma_D=3.4\,\micron$.} \label{fig:powerlaw}
\end{figure}

\begin{figure}
\includegraphics[angle=-90,width=1.0\columnwidth]{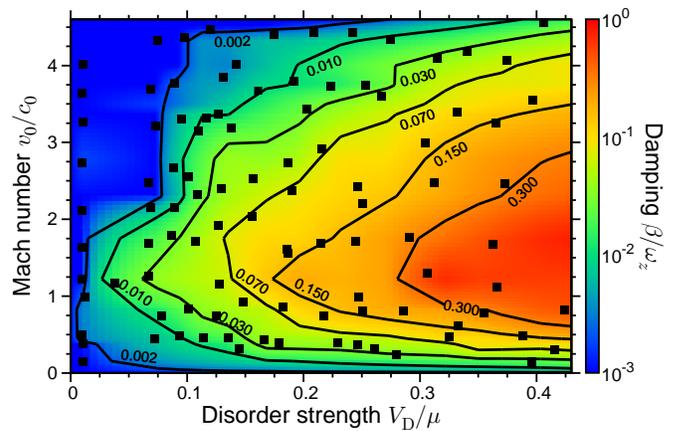}
\caption{(Color) Transport regimes of a BEC traveling through a disordered harmonic potential. Black squares show
the values of disorder strength $\Vd/\mu$ and initial peak center of mass velocity $v_0/c_0$ for the data used to
extract $\beta$ from a fit to Eq.~\ref{eqn:osc} using 4--6 periods of oscillation. The interpolated color map
(and contour lines) for $\beta/\omega_z$ is derived from the measured results. These measurements have
$a=200\,a_0$, $N=2\times10^5$ atoms, $\mu / h=1.5\,\mathrm{kHz}$, $c_0=6.5$\,mm/s$, \omega_r =
(2\pi)\,260\,\mbox{Hz}$, and $\omega_z = (2\pi)\,5.5\,\mbox{Hz}$. 
The variable experimental quantities are $A$ and $\Vd$.
Due to small shot-to-shot fluctuations in the
position of the center of mass of the cloud, measurements with $v_0 < 0.2\,c_0$ are not reliable. Data with
$\beta \leq 2 \times 10^{-3}$ is consistent with undamped motion.} \label{fig:phasedis}
\end{figure}
We have systematically investigated the dependence of $\beta$ on the
disorder strength $\Vd$.  Figure \ref{fig:powerlaw} shows the
normalized damping parameter $\beta/\omega_z$ plotted against the
normalized disorder strength $\Vd / \mu$, where $\mu$ is the
chemical potential of the condensate prior to the kick and before
the disorder is switched on.  We find the data fits well to a power law
\begin{equation}\label{eqn:powerlaw}
\frac{\beta}{\omega_z} \propto \left(\frac{\Vd}{\mu}\right)^{p},
\end{equation}
\noindent for all measured velocities. The precise value of $p$,
however,  depends weakly on $v_0$ across the range of velocities $0
< v_0/c_0 < 5$,  with a mean value of $p=2.1(5)$ (see Fig.~\ref{fig:powerlaw} inset).

Figure~\ref{fig:phasedis} presents the measured values of $\beta$ as
a function of both $\Vd$ and $v_0$. {As expected, a
vertical trace through this plot shows a qualitative similarity to
Fig.~\ref{fig:vdamp}.} We observe two distinct regimes of reduced
damping: one where $v_0 / c_0 \ll 1$ and the other when $v_0 / c_0
\gg 1$, with the damping reaching a maximum at $v_0 \sim c_0$.
A numerical simulation using an effective 1D NLSE has
produced qualitatively similar results \cite{albert08}.

\subsection{Variation with Interaction Strength}

\begin{figure}
\includegraphics[angle=-90,width=1.0\columnwidth]{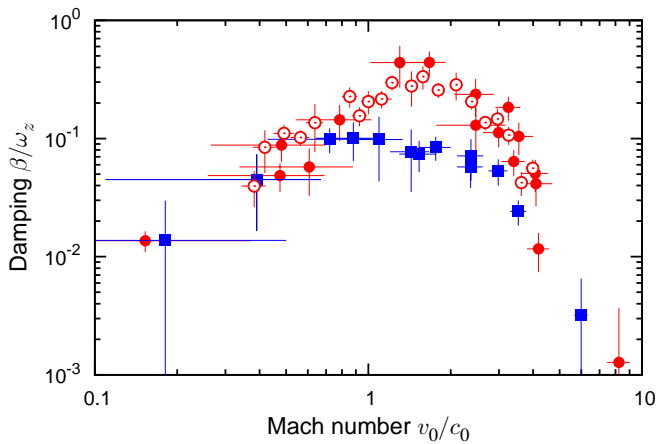}
\caption{(Color online) Universal damping vs.~$v_0/c_0$. The disorder strength was adjusted to keep $0.30<\Vd/\mu<0.35$ for all
of the data.  Squares correspond to $a=28\,a_0$, $N=2.5\times10^5$, $\mu / h = 550\,$Hz, $c_0 = 4.0\,$mm/s,
$\omega_z=(2\pi)\,5.5\,$Hz, and $\omega_r=(2\pi)\,260\,$Hz; open circles correspond to $a=200\,a_0$,
$N=3\times10^5$, $\mu / h = 2.4\,$kHz, $c_0 = 8.3\,$mm/s, $\omega_z=(2\pi)\,4.5$\,Hz, and
$\omega_r=(2\pi)\,460$\,Hz; filled circles correspond to the same parameters as Fig.~\ref{fig:phasedis}. Error
bars are as defined in Fig.~\ref{fig:vdamp}. } \label{fig:damping}
\end{figure}

We observe nearly universal behavior for $\beta$ as a function of both $\Vd/\mu$, and $v_0/c_0$ for BECs in the
Thomas-Fermi regime. As already shown in Fig.~\ref{fig:powerlaw}, $\beta\propto\left(\Vd/\mu\right)^2$
for condensates with $\mu$ differing by a factor of 3. Shown in
Fig.~\ref{fig:damping} is a comparison between the damping at interaction strengths $a=200\,a_0$ and $a=28\,a_0$,
with constant $\Vd/\mu$. Although the respective values of $c_0$ differ by nearly a factor of 2 between the two
data sets, the peak damping occurs at $v_0/c_0\sim1$ for both, demonstrating the nearly universal behavior of
$\beta$ vs.~$v_0/c_0$.  On the other hand, the peak damping rate between the two data sets differs by nearly a
factor of 5, showing that while the general shape of the damping curve is universal, the magnitude of the damping
is not.

\begin{figure}
\includegraphics[angle=-90,width=1.0\columnwidth]{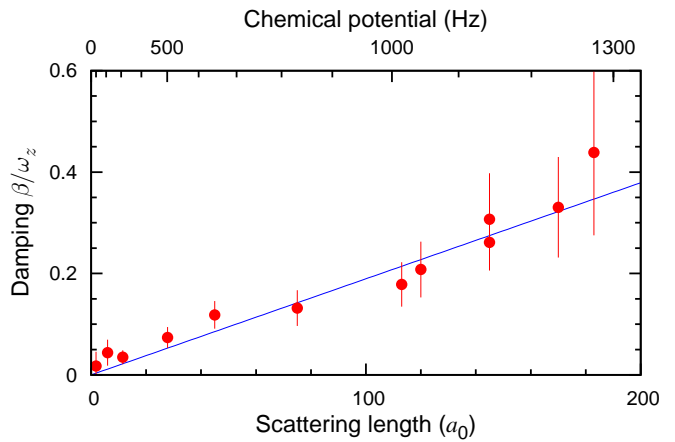}
\caption{(Color online) Peak damping vs.~$a$ with fixed $\Vd/\mu$ and $v_0/c_0$.  For this data, $\Vd$ and $v_0$ were adjusted
to keep $0.3<\Vd/\mu <0.4$ and $0.6<v_0/c_0<1.4$ with all other parameters as in Fig.~\ref{fig:phasedis}.  The
upper horizontal axis shows values for $\mu$ obtained from a variational solution of the GPE \cite{pollack09} (note
 that the upper tick marks are not strictly logarithmically spaced). 
The linear fit has a slope $0.002\,a_0^{-1}$.
Vertical error bars are as defined in Fig.~\ref{fig:vdamp}. }
 \label{fig:varya}
\end{figure}

An investigation of the effect of interatomic interactions on the peak
damping ($v_0/c_0\sim1$) at fixed $\Vd/\mu$ is shown in
Fig.~\ref{fig:varya}.  We find that $\beta$ scales linearly with
$a$, going to zero with decreasing interactions, consistent with the
disappearance of the low energy phonon portion of the excitation
spectrum as $U\rightarrow0$.

\begin{figure}
\includegraphics[angle=-90,width=1.0\columnwidth]{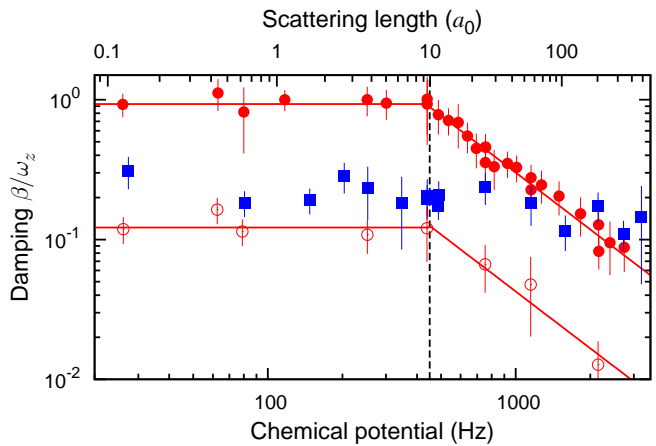}
\caption{(Color online) Damping vs.~$\mu$ with fixed $\Vd$ and $v_0$. Squares and filled circles correspond to $\Vd /h=370\,$Hz
with $v_0 = 11\,$mm/s and $6\,$mm/s, respectively. Open circles correspond to $\Vd / h = 140\,$Hz and $v_0 =
6\,$mm/s. The vertical dashed line denotes $\mu = \hbar \omega_r$,
at which point $v_0/c_0 = 1.7$ for the open and filled circles and $v_0/c_0 = 3$ for the squares.
We varied $\mu$ by adjusting $a$, shown on the
upper horizontal axis (note that the upper tick marks are not strictly logarithmically spaced). Values for $\mu$
are obtained from a variational solution of the GPE \cite{pollack09} using the following measured experimental
parameters: $\omega_r=(2\pi)\,460$\,Hz, $\omega_z=(2\pi)\,4.5$\,Hz, and $N=4 \times 10^5$ atoms. For this data
$\sigma_D = 3.4\,\micron$. Vertical error bars are as defined in Fig.~\ref{fig:vdamp}. } \label{fig:dampva.ps}
\end{figure}
The elongated confinement geometry in our system facilitates the
investigation of the dimensional crossover from the 3D to the
quasi-1D regime where $\mu \ll \hbar\omega_r$
\cite{gorlitz01,Schreck01}. Shown in
Fig.~\ref{fig:dampva.ps} are measurements
of $\beta$ vs.~$\mu$ at constant $\Vd$ and $v_0$.
When $\mu>\hbar\omega_r$ (to the right of the vertical dashed line) and $v_0$ is comparable to, or less than $c_0$ (as is the case for the data shown as open and closed circles),
we find $\beta\propto\mu^{-1.4}$.
By reference to Fig.~\ref{fig:phasedis}, one can gain a
qualitative understanding of this behavior going from high to low $\mu$:
starting subsonically (open and filled circles),
the system travels
along a path from the weakly damped regime (lower left corner of Fig.~\ref{fig:phasedis})
towards the regime of strong damping (middle right region).  As $\mu$ decreases, the quantities $\Vd/\mu$ and $v_0/c_0$ increase correspondingly, and the system follows a path which crosses several contours of constant $\beta$ while approaching the strongly damped regime near $v/c\sim1$. Consequently, the system displays a strong dependence of $\beta$ on $\mu$. 
Blue squares depict a different situation where the system is supersonic for all $\mu$ investigated.  For large $\mu$ the system occupies a point in Fig.~\ref{fig:phasedis} with $v_0/c_0>1$ and $\Vd/\mu<1$.  As $\mu$ decreases, the system follows a diagonal path, roughly tracing a contour line of constant $\beta$, moving into the regime of $v_0/c_0\gg1$ and $\Vd/\mu\gg1$ (top-right corner of Fig.~\ref{fig:phasedis}).    
When $\mu < \hbar\omega_r$ (to the left of the vertical dashed line), and $v_0 \gg c_0$,
we observe a negligible dependence of $\beta$ on $\mu$.
In this quasi-1D regime,
$\beta$ is affected only by changing $\Vd$ or $v_0$, consistent with the
behavior expected for a nearly ideal, classical fluid.
This may be understood by reference to Eq.~\ref{eqn:spec}
where for $v\gg c$, the first term in the Bogoliubov excitation spectrum
dominates making the system ``quasi-ideal'' with $\epsilon(p)$ independent of $\mu$.
\begin{figure}
\includegraphics[angle=-90,width=1.0\columnwidth]{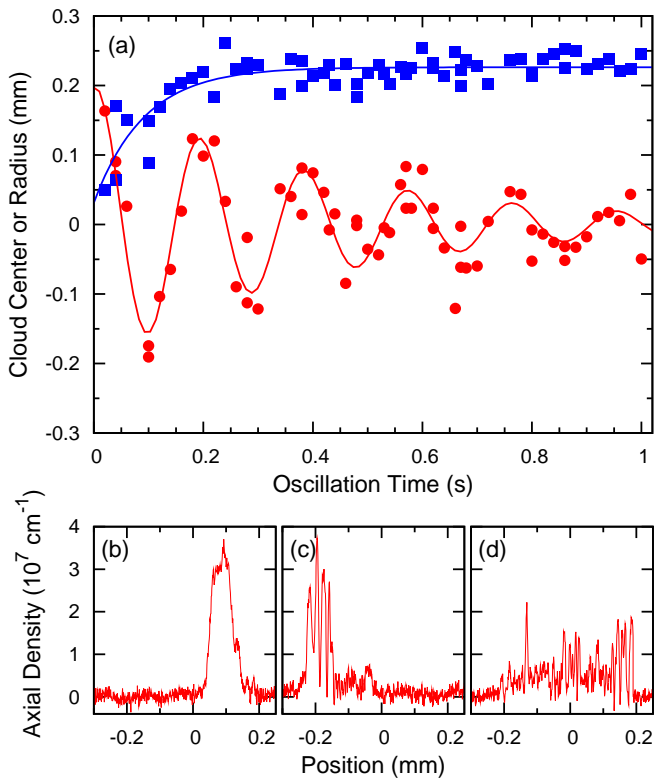}
\caption{(Color online) Damping of a nearly non-interacting gas. For this data $a = 0.4\,a_0$, $N=2\times10^5$, $\mu / h =
26\,\mathrm{Hz}$, $c=1.2\,$mm/s, $\Vd = 4\,\mu$, $\omega_r=(2\pi)\,240$\,Hz, and $\omega_z=(2\pi)\,5.3$\,Hz. (a)
Center of mass position (circles) and radius (squares) of the condensate as a function of time. Here we use
statistically determined values for the center of mass $z_\mathrm{cm} = \int z\, n(z) dz / N$ and radius $R$,
given by $R^2 =  4 \int (z - z_\mathrm{cm})^2\, n(z) dz / N$. (b--d) Axial density traces at various times in the
oscillation: (b) 40\,ms (c) 100\,ms, (d) 960\,ms. After two full oscillations, the cloud has fragmented and
spread to a size comparable with the initial oscillation amplitude.}\label{fig:weakdamp_wtrace}
\end{figure}

Figure~\ref{fig:weakdamp_wtrace} shows damping of a weakly interacting gas with $a = 0.4\,a_0$, deep into the
quasi-1D regime, where $\mu / \hbar \omega_r \sim 0.1$. We find that $\Vd=4\,\mu$ produces the same damping
($\beta/\omega_z=0.07$) as that for a BEC with $a = 200\,a_0$ and $\Vd=0.25\,\mu$. The nature of the damped
motion of a weakly interacting gas in strong disorder is strikingly different from the damped motion of a
strongly interacting gas in weak disorder, even though the timescale of the damping in both cases is comparable.
Figure~\ref{fig:weakdamp_wtrace} shows that the damping in the weakly interacting regime is caused by the loss of
coherence of the collective dipole mode brought on by extensive fragmentation. Because $\Vd > \mu$, it is perhaps
not surprising that the condensate quickly fragments. While the center of mass of the cloud damps after about 5
oscillation periods, examination of shot-to-shot differences in the damped density distributions reveal that the
position of the fragments are highly non-repeatable, suggesting that some fragments remain in motion. This
residual motion is consistent with the long thermalization time expected from weak two-body interactions. It is
interesting to note that the maximum single particle kinetic energy, $E_\mathrm{K} = \frac{1}{2} m \omega_z^2 A^2
= h\,(295\,\mathrm{Hz})$, is 2.8 times larger than the average height of the disordered potential. The observed
dephasing is therefore consistent with the expected behavior of a gas of non-interacting particles interacting
with a disordered potential where the disorder strength is smaller than the kinetic energy of the individual
particles.

\section{Dissipation Induced by a Single Gaussian Defect}\label{sec_bar}
\subsection{Thomas-Fermi Regime}
\begin{figure}
\includegraphics[angle=-90,width=1.0\columnwidth]{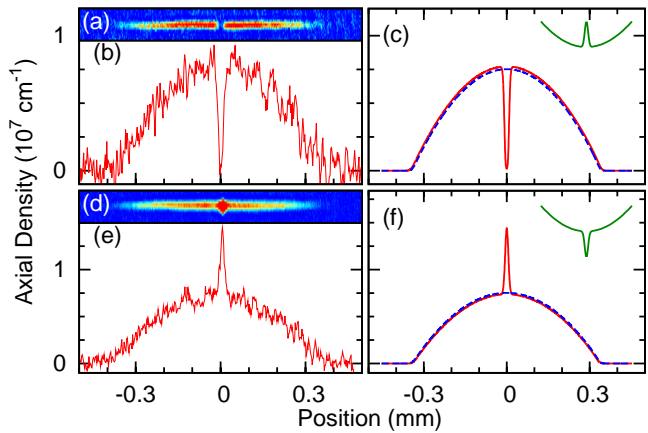}
\caption{(Color online) BEC in a harmonic trap with a single Gaussian defect. (a--c) correspond to a repulsive defect, while
(d--f) to an attractive one. (a) and (d) \textit{in situ} polarization phase-contrast images, (b) and (e) axial
densities corresponding to the images, and (c) and (f) numerical solutions to the GPE with the dashed lines
showing the solution in absence of a defect. The inset trace shows the characteristic shape of the potential. For
all panels, $a=200\,a_0$, $N=4\times10^5$, $\omega_z=(2\pi)\,5.0\,$Hz, and $\omega_r=(2\pi)\,360\,$Hz.} \label{fig:impurities}
\end{figure}

In an effort to better understand the mechanisms responsible for the damping by disorder, we have
investigated the dissipation induced by a \textit{single} Gaussian defect. The defect potential is described by
$V(z) = \Vd e^{-2 z^2/w_z^2}$, where $w_z = 12\,\micron$. The static effect of either an attractive or repulsive
defect on a repulsively interacting BEC in the Thomas-Fermi regime is shown in Fig.~\ref{fig:impurities}.  As
expected, the attractive defect leads to an increase of the density in the region of the defect, accompanied by a
small decrease of the density in the wings of the distribution, while the opposite is true for a
repulsive defect.
\begin{figure}
\includegraphics[angle=-90,width=1.0\columnwidth]{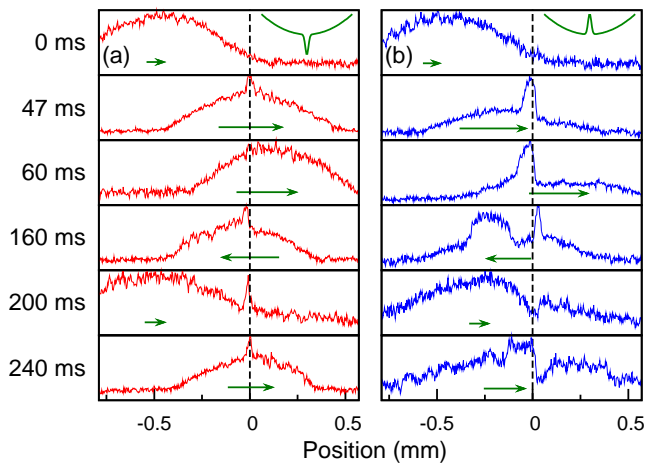}
\caption{(Color online) Axial densities at various times during a supersonic oscillation in the presence of an (a) attractive or
(b) repulsive defect.
The arrows are proportional to the instantaneous
velocity of the condensate. 
The vertical dashed lines denote location of the defect. 
For this data, $a=200\,a_0$ and $v_0=13\,$mm/s;
(a) corresponds to
$N=4\times10^5$, $\mu=1.5\,$kHz, $v_0/c_0=2$, $\Vd=-0.8\,\mu$, $\omega_z=(2\pi)\,4.7\,$Hz, and
$\omega_r=(2\pi)\,360\,$Hz; 
(b) corresponds to $N=1\times10^6$, $\mu=3\,$kHz, $v_0/c_0=1.4$, $\Vd=0.4\,\mu$,
$\omega_z=(2\pi)\,5.0\,$Hz, and $\omega_r=(2\pi)\,360\,$Hz.} \label{fig:dampedosc_insitu}
\end{figure}
\begin{figure}
\includegraphics[angle=-90,width=1.0\columnwidth]{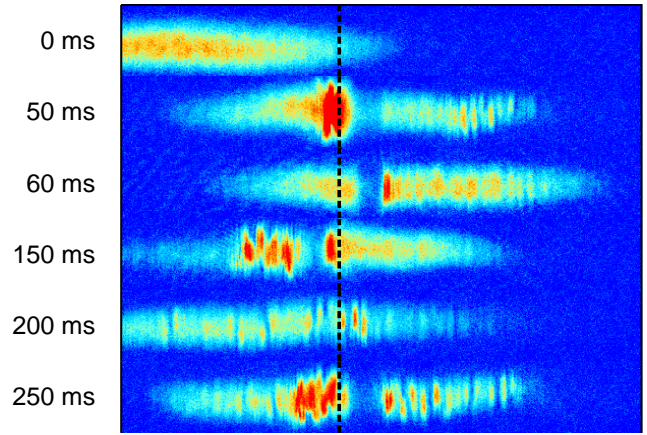}
\caption{(Color online) Density fluctuations produced crossing a repulsive defect. Absorption images after 4\,ms time-of-flight.
The vertical dashed line denotes the location of the defect.  The experimental parameters are as stated in
Fig.~\ref{fig:dampedosc_insitu}(b).} \label{fig:dampedosc}
\end{figure}

The dynamical distributions can differ dramatically from the static
case, as shown in Fig.~\ref{fig:dampedosc_insitu} where \textit{in situ}
axial densities are displayed for various times throughout the
dipole oscillation.
In the following discussion we refer to the
upstream side of the condensate as the portion of the
BEC that reaches the barrier after the leading or downstream portion.
The interaction of the BEC with the repulsive defect, shown in
Fig.~\ref{fig:dampedosc_insitu}(b), produces a deep downstream
density rarefication as well as a large upstream density compression
bearing a qualitative similarity to a shock wave. Similar structures
have been predicted in effective 1D theoretical treatments
\cite{Leszczyszyn09,salasnich02} and interpreted as upstream and
downstream dispersive shocks.  In contrast, the interaction of the
BEC with the attractive defect, shown in
Fig.~\ref{fig:dampedosc_insitu}(a), produces no such shock waves.
However, the cloud is slightly compressed near the
defect simply due to the attractive defect potential.  Because
$v_0>c_0$, phonon excitations cannot be emitted in the upstream
direction as they would have to propagate faster than the speed of
sound.  Close inspection of Fig.~\ref{fig:dampedosc_insitu}(a)
reveals minimal density modulation of the upstream side, while
more modulation is evident on the downstream side.

Several 1D theoretical studies predict the formation of downstream
propagating dark solitons in addition to an upstream dispersive shock
as a repulsive defect is supersonically swept through a condensate
\cite{leboeuf01,pavloff02,Radouani04,Theocharis05,Carretero07,albert08,salasnich02,Leszczyszyn09},
which is consistent with the density fluctuations visible in
Fig.~\ref{fig:dampedosc_insitu}. 
However, the size of the dark
solitons will be on the order of the healing length,
$\xi=0.5\,\mu$m for these condensates, which is a factor of 6
smaller than our imaging resolution.

Figure~\ref{fig:dampedosc} shows time-of-flight
images of the BEC oscillating in the presence of a single repulsive defect.
In contrast to the
\textit{in situ} images of Fig.~\ref{fig:dampedosc_insitu}(b),
after time-of-flight additional structures
emerge which were not previously visible.
These structures are consistent with dark solitons that form from
short length scale \textit{in situ} phase fluctuations that map onto larger scale density modulations after time-of-flight.
However, \textit{in situ} phase fluctuations may also arise from thermal
excitations in highly elongated BECs,
and these can also manifest as density fluctuations after time-of-flight \cite{dettmer01}.
Close inspection of Fig.~\ref{fig:dampedosc} reveals that
deep density modulations are present only in
the downstream portion of the BEC (after the first pass through the defect),
consistent with the dark soliton interpretation.
Similar density fluctuations have also been
interpreted as dark solitons in an experiment using a moving defect
and a stationary BEC~\cite{engels07}.

\begin{figure}
\includegraphics[angle=-90,width=1.0\columnwidth]{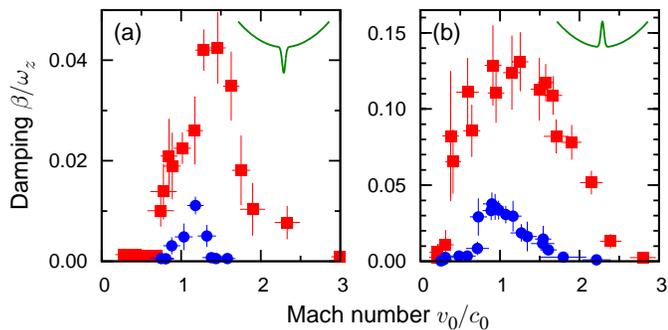}
\caption{(Color online) Velocity dependence of $\beta$ induced by a single Gaussian defect. 
(a) Attractive defect with (squares) $\Vd/\mu = -0.8$ or (circles) $\Vd/\mu = -0.3$
and other parameters as stated in Fig.~\ref{fig:dampedosc_insitu}(a) except with $N=8\times10^5$ and $\mu/h=2\,$kHz. 
(b) Repulsive defect with (squares) $\Vd/\mu = 0.4$ or (circles) $\Vd/\mu = 0.2$
and other parameters as stated in Fig.~\ref{fig:dampedosc_insitu}(b). 
Both types of impurities show critical behavior at low velocities as well as
undamped motion at large $v_0/c_0$. Note the difference in scale between damping induced by an attractive versus
a repulsive impurity.  Vertical and horizontal error bars are as described in Fig.~\ref{fig:vdamp}.
}\label{fig:bluebar}
\end{figure}

\begin{figure}
\includegraphics[angle=-90,width=1.0\columnwidth]{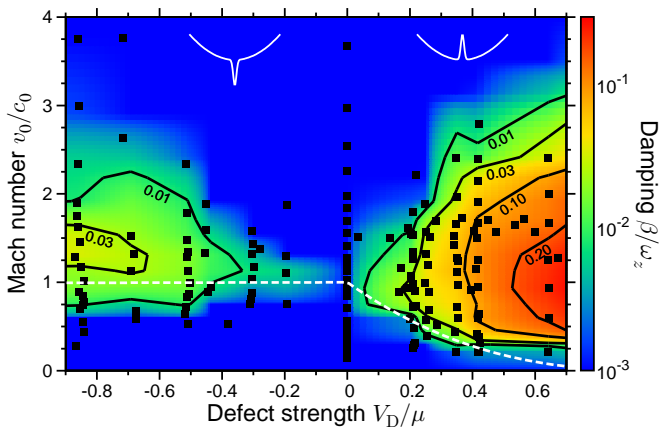}
\caption{(Color) Transport regimes of a BEC traveling through a harmonic potential with a central Gaussian
defect. Coordinates of the black squares are the values of $\Vd/\mu$ and $v_0/c_0$ for the data sets used to
extract $\beta/\omega_z$ from a fit to Eq.~\ref{eqn:osc}. The color map (and contour lines) for $\beta/\omega_z$
is derived from an interpolation using the measured results. Dashed white lines show the local Landau critical
velocity as given by Eqs.~\ref{eqn:vcbar} and~\ref{eqn:redvc}.  The attractive and repulsive cases are
qualitatively similar: superfluidity for $v_0/c_0\ll1$, increased damping as $v_0/c_0\rightarrow 1$, and reduced
damping for $v_0/c_0\gg 1$. Damping induced by an attractive impurity is an order of magnitude weaker than for a
repulsive one.  Data with $\Vd<0$ and $\Vd>0$ correspond to parameters in Figs.~\ref{fig:dampedosc_insitu}(a) and
(b), respectively.} \label{fig:bardamp}
\end{figure}

We have measured $\beta$ as a function of both $\Vd$ and $v_0$, with characteristic results presented in
Fig.~\ref{fig:bluebar}. Contrary to what was observed for a disordered potential, we observe a critical velocity
$v_c$ below which the motion is undamped, for both the attractive and repulsive defects. We find that the peak
damping for an attractive defect is significantly weaker than for a repulsive one. 
Figure~\ref{fig:bardamp} presents measurement results 
of $\beta$ as a function of both $\Vd$ and $v_0$.
For an attractive defect, we find that $v_c/c_0 \sim 0.7$ with $v_c$ depending only weakly on $\Vd$. 
However, for a moderately strong repulsive defect, $v_c/c_0$ occurs significantly below $1$ and depends strongly on $\Vd$. For
both attractive and repulsive defects $v_c$ tends to $c_0$ as $|\Vd/\mu|$ is reduced to zero.

Once again, a model based on a local Landau criterion is
sufficient to explain the dependence of $v_c$ on $\Vd$. For simplicity,
consider a uniform density flow impinging on either a repulsive or
attractive Gaussian potential \cite{pavloff02}. With the assumption
that the superfluid flow pattern is stationary, the local density of
the condensate near the defect must be modified in a similar way to
that shown in Fig.~\ref{fig:impurities} for a static defect. For the
repulsive case, the local density is reduced
near the defect, resulting in a lower local
speed of sound. In addition, flux conservation requires that the
local condensate velocity increase in the low 
density region near the repulsive impurity to preserve
the stationary flow pattern. 
A corresponding argument can be made for the case of an attractive defect.
These effects serve to increase
the local value of $v(z)/c(z)$ near a repulsive defect and decrease it for an
attractive one. As a result, excitations can be created near the
peak of the repulsive defect in a BEC with a center of mass velocity that is
significantly lower than the bulk speed of sound. For the case of an
attractive impurity, on the other hand, one expects excitations to
occur in the bulk condensate first, rather than near the impurity,
and therefore at a flow velocity near the bulk speed of sound, as
observed.

We quantify this picture, in the case of a repulsive defect, by
applying the local Landau criterion at the instant the center of the
BEC crosses the defect.  
In Ref.~\cite{albert08} the authors 
used an effective 1D NLSE in the high
density regime to determine the locus of points where the local condensate
velocity $v(z)$ is equal to the local speed of sound $c(z)$; this defines
the curve
\begin{eqnarray}
\frac{v_c}{c_0} = \left( 1 - \frac{\Vd}{\mu}\right)^{5/2}, & \Vd>0 \label{eqn:vcbar},
\end{eqnarray}
\noindent where $\Vd/\mu \equiv \delta n_0/n_0$ is the fractional change in
the peak density at the peak of the repulsive defect.
When $v_0/c_0<1$, we can ignore effects of the axial Thomas-Fermi
profile of the condensate because $A\ll R_\mathrm{TF}$ for our trap,
where $R_\mathrm{TF}$ is the axial Thomas-Fermi radius.
Equation~\ref{eqn:vcbar} is plotted in Fig.~\ref{fig:bardamp} when $\Vd>0$ and
is found to agree with the measured $v_c$ for the range
of $\Vd$ explored experimentally. Therefore, the observed reduction
of the critical velocity below $c_0$ is consistent with the \textit{local}
Landau critical velocity without invoking more exotic mechanisms,
such as vortex nucleation.  This is in contrast with
several experiments involving BECs in less elongated configurations
\cite{onofrio00,Raman01}, as well as in superfluid $^4$He where nucleation
of vortex lines and rings can result in $v_c<v_L$~\cite{tilleybook}.

In the case of the attractive defect, the density, and therefore
$c(z)$, is enhanced at the location of the defect and reduced only
slightly elsewhere. We find that the reduction in density in the
bulk due to the enhancement at the defect is less than $1\%$ for the
strongest barriers used, leading to an essentially unperturbed speed
of sound in the bulk.  The ratio of the local fluid velocity to the
local speed of sound can then be found by considering only the bare
Thomas-Fermi profile, and is given by
\begin{equation}
\frac{v(z)}{c(z)}=\frac{v_0}{c_0}\left(\frac{1-z^2/A^2}{1-z^2/R_\mathrm{TF}^2}\right)^{1/2},
\end{equation}
where, using Eq.~\ref{eqn:sound}, $v_0/c_0=2A/R_\mathrm{TF}$. If $2 A < R_\mathrm{TF}$ then
$v_0/c_0<1$ and the local Landau criterion is satisfied everywhere inside the condensate:
\begin{eqnarray}\label{eqn:redvc}
\frac{v_c}{c_0} = 1, & \Vd<0,
\end{eqnarray}
implying that $v_c$ is independent of $\Vd$.  Our measurements, however, show that $v_c$ depends weakly 
on $\Vd$ with $v_c/c_0\rightarrow1$ only in the weak impurity limit. 
Our experimental results are consistent with numerical simulations using a 1D NLSE \cite{albert08} for which the local Landau criterion
accurately describes the repulsive impurity case, but slightly overestimates $v_c$ in the attractive case.

\begin{figure}
\includegraphics[angle=-90,width=1.0\columnwidth]{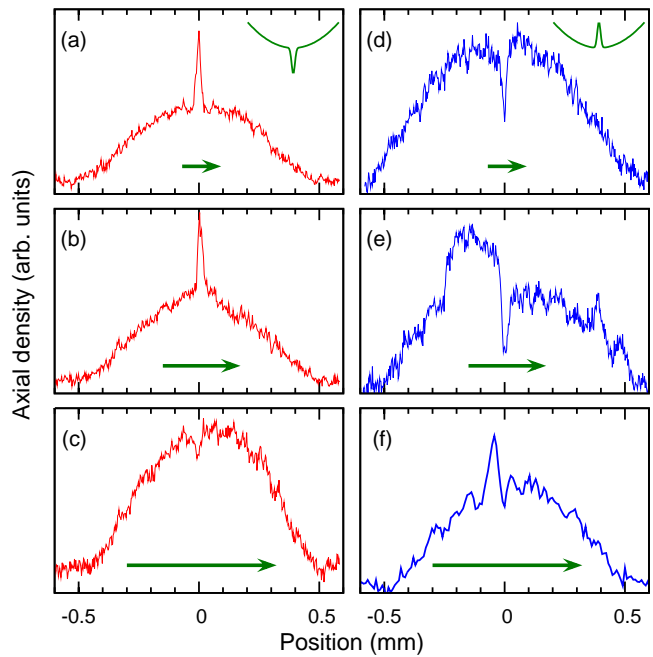}
\caption{(Color online) \textit{In situ} density distributions of a condensate passing through a Gaussian defect. These images
are taken at the instant the center of the BEC first crosses the defect. Rows correspond to the three flow
regimes: subsonic superfluid ($v_0/c_0<1$), dissipative ($v_0/c_0\sim1$), and supersonic non-dissipative
($v_0/c_0>1$). (a--c) Attractive defect with $\Vd/\mu=-0.85$ and $v_0/c_0=0.31$, $v_0/c_0=1.0$, and
$v_0/c_0=3.0$, respectively. (d--f) Repulsive defect with $\Vd/\mu=0.65$ and $v_0/c_0=0.15$, $v_0/c_0=0.90$, and
$v_0/c_0=2.0$, respectively.  
The arrows indicate the direction and relative speed of the condensate.
For this data, all other parameters are as described in Fig.~\ref{fig:dampedosc_insitu}.
}\label{fig:combo}
\end{figure}

Figures~\ref{fig:bluebar} and \ref{fig:bardamp} demonstrate that damping is significantly suppressed deep
into the supersonic regime. We observe undamped motion when $v_0$ is greater than a $\Vd$-dependent upper
critical velocity $v_+$. Numerical simulations \cite{hakim97,Radouani04,albert08} have shown that for ``wide and
smooth'' barriers ($\xi \ll w_z$) the emission of radiation from the defect in the form of phonons and solitons
can be very small for supersonic velocities. In fact, it has been shown analytically that the radiation emission
rate resulting from a defect moving supersonically through a condensate decreases exponentially with the ratio
$\xi/w_z$ \cite{haddad01}. Without emission of radiation, energy dissipation is inhibited and the flow persists,
even though Landau's criterion is violated. For the data presented in
Figs.~\ref{fig:impurities}--\ref{fig:combo}, $\xi/w_z \sim 0.04$, well within the regime where supersonic
non-dissipative flow is predicted. Experiments similar to ours have also shown a reduction in soliton emission
from a barrier moving through a condensate in the supersonic regime \cite{engels07}.

We therefore observe three distinct regimes of flow in the single
defect system: subsonic superfluid ($v_0/c_0<1$), dissipative
($v_0/c_0\sim 1$), and supersonic non-dissipative
{($v_0/c_0>v_+$).} Figure~\ref{fig:combo} displays 
axial densities from \textit{in situ} polarization phase-contrast
images at the instant the defect passes through the peak of the
condensate for the three different velocity regimes. As expected,
for the superfluid flow regime the axial density profiles look very
much like the equilibrium profiles of Fig.~\ref{fig:impurities}:
there is an increase (decrease) in the density at the location of
the attractive (repulsive) defect. In the dissipative flow regime,
on the other hand, the flow patterns for $\Vd>0$ show
significant distortion, while for $\Vd<0$ there is little
distortion, as discussed in detail above.
Finally, in the supersonic non-dissipative flow regime, we observe a
counter-intuitive density \textit{inversion} with respect to the
superfluid regime, where the attractive defect produces
a density depression while the repulsive defect causes a density peak.

The physical origin of this counter-intuitive
density inversion can be understood by considering the behavior of
the gas at large $v_0$.  In this regime, as in the disordered case,
the Bogoliubov excitation spectrum, given by Eq.~\ref{eqn:spec}, is
dominated by the $p^2/2m$ term
and therefore, dominated by plane waves
with wavenumber $k = p / \hbar$ rather than phonons.  
For this ``quasi-ideal'' gas, the
drag should be determined by the scattering of these plane waves off of
the defect \cite{pavloff02}.  
At high velocities, scattering of these waves from the defect is greatly
suppressed, leading to low dissipation.
If we extend this argument further and consider the
atoms to be classical particles, one expects the atoms to slow
down in the presence of the repulsive defect,
resulting in a density increase near the defect,
while the opposite is expected for an attractive defect.

Density inversions similar to the ones presented here have also been discussed in the context of dissipationless
stationary states at supersonic velocities \cite{law00,haddad01,law01,leboeuf01} as well as sonic black holes \cite{lahav09}.  
Under our experimental conditions, when $v_0/c_0\sim1$ the edge of the barrier can serve as a sonic event horizon. 
Such systems have been proposed as possible candidates with which to study ``table top'' astrophysics, where exotic
effects, such as Hawking radiation, should be observable. Interestingly, in this system the experimenter plays
the role of the so-called super-observer, having access to the regions both outside and \textit{inside} the event
horizon~\cite{garay00,balbinot08,carusotto08,lahav09}.

\subsection{Weakly Interacting Regime\label{sec:weakbar}}

\begin{figure}
\includegraphics[angle=-90,width=1.0\columnwidth]{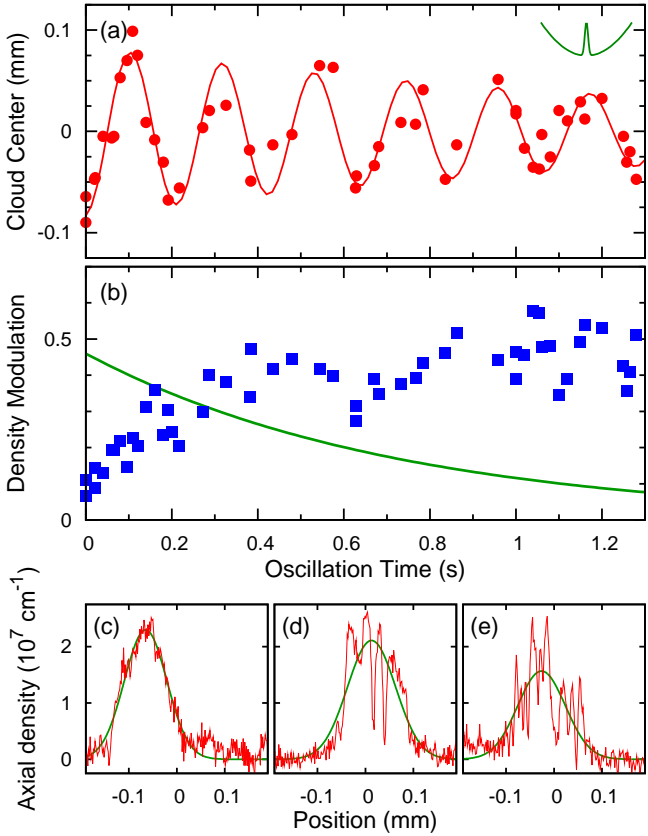}
\caption{(Color online) Oscillation of a weakly interacting BEC in the presence of a repulsive defect, with $a = 0.6\,a_0$,
$N=2.5\times10^5$, $\mu / h =44\,$Hz, $\Vd/\mu = 0.8$, $\omega_z= (2\pi)\,4.7\,$Hz, $\omega_r=(2\pi)\,300\,$Hz,
and $v_0/c_0=1.6$.
(a) Center of mass position as a function of time (computed as in
Fig.~\ref{fig:weakdamp_wtrace}) giving $\beta/\omega_z=0.03$; 
(b) Root-mean-square density deviation from a Gaussian fit to the axial density distribution (see text).
The solid line shows the decay of the oscillation energy (in arbitrary units) found from the fit in (a). 
(c--e) \textit{In situ} axial density traces and Gaussian fits at various oscillation 
times: (c) 0\,ms; (d) 140\,ms, at the second crossing of the defect; and
(e) 1260\,ms, after several crossings of the defect. At large times we find that the large density modulations
are accompanied by only a slight increase of the axial size of the condensate. } \label{fig:weakdamposc}
\end{figure}
Figure~\ref{fig:weakdamposc} shows results of measurements of a weakly interacting condensate ($a = 0.6\,a_0$) oscillating in
the presence of a repulsive defect.  Under these conditions the condensate is in the quasi-1D regime with
$\mu/\hbar\omega_r = 0.15$. We find that the axial density profile of the condensate becomes increasingly
modulated during the damped oscillation, consistent with theory~\cite{Radouani04,Theocharis05}. 
We compute the root-mean-square deviation of the axial density distribution from a Gaussian fit $n_\mathrm{fit}$
as a proxy for the increased internal energy of the condensate due to the density modulation
\begin{equation}
\Delta = \sqrt{ \frac{1}{L} \int_L \left[
    \frac{ n(z) - n_\mathrm{fit}(z)}{n_\mathrm{fit}(z)}
\right]^2 dz },
\end{equation}
\noindent where we take the integration length $L$ to be over the
central 70\% of the condensate to minimize edge effects.
Figure~\ref{fig:weakdamposc}(b) shows that $\Delta$
initially increases in time and then saturates.
The time-dependent increase in the density modulation 
qualitatively matches the loss of oscillation energy. 
We therefore conclude that the damping of the dipole mode is caused
primarily by the creation of \textit{in situ} density modulations in
the cloud.

\begin{figure}
\includegraphics[angle=-90,width=1\columnwidth]{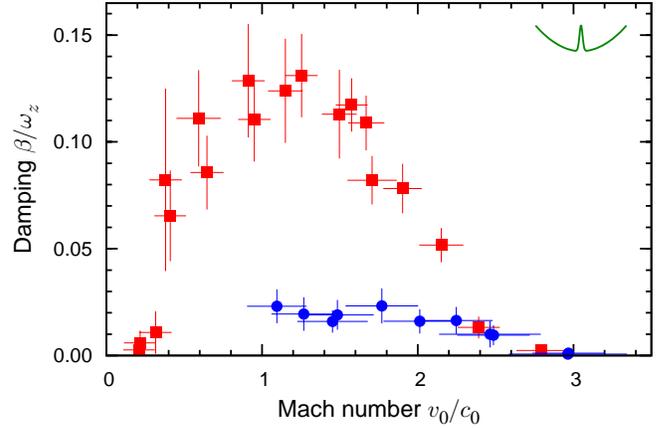}
\caption{(Color online) Velocity dependence of damping with a repulsive defect.
Circles correspond to a nearly non-interacting BEC
    with $a=0.6\,a_0$, $\mu / h =31$\,Hz, $c_0 = 0.9\,$mm/s, and $\Vd=0.9\,\mu$.
Shot-to-shot variations in the position of the cloud limit the
extraction of $\beta$ to $v_0 > 1\,$mm/s, corresponding to $v_0/c_0
> 1.1$. Squares correspond to data from Fig.~\ref{fig:bluebar}(b)
for comparison, $a = 200\,a_0$, $\mu / h = 3\,$kHz, $c_0 =
9.2\,$mm/s, and $\Vd = 0.4\,\mu$.  Error bars are as defined in
Fig.~\ref{fig:vdamp}. } \label{fig:weakdamp}
\end{figure}

Results of measurements of the velocity-dependence of
the damping by a repulsive defect with $a=0.6\,a_0$ are shown in Fig.~\ref{fig:weakdamp}.
As was the case with disordered potentials, we find
that the timescale for damping in the quasi-1D regime
with a strong impurity strength
is much longer than that observed in the Thomas-Fermi regime
with a weak impurity strength.

\subsection{Dark Soliton Production in the Weakly Interacting Regime\label{sec:solitons}}

Of particular interest in the quasi-1D regime is the ability to
create and observe long-lived dark solitons. These
nonlinear excitations have been previously created in BECs with
repulsive interatomic interactions through a variety of means,
including direct phase imprinting \cite{burger99,Denschlag2000},
spatially selective microwave transfer \cite{anderson01}, slow light
\cite{dutton01}, two condensate interference
\cite{weller08,chang08}, and, similar to the work presented here, as
a result of a BEC crossing a semi-permeable defect \cite{engels07}.

In general, the decay of dark solitons occurs as a result of
dynamical instability or as a result of dissipative dynamics
associated with the interaction of the soliton with quasiparticle
excitations of the BEC. {However, it is known that
dark solitons can have very long lifetimes in the quasi-1D regime
\cite{Muryshev02}.} For the most weakly interacting BECs presented
here, $\mu/\hbar\omega_r = 0.13$, making our system ideally suited
to study long-lived dark solitons.
\begin{figure}
\includegraphics[angle=-90,width=1\columnwidth]{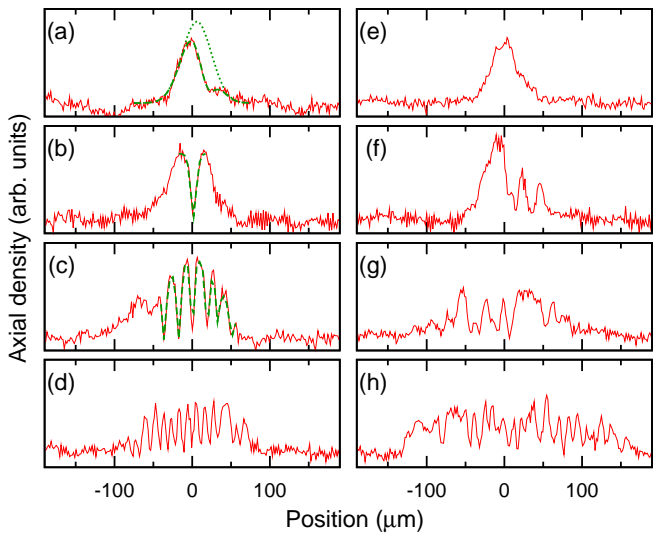}
\caption{(Color online) Dark soliton formation. \textit{In situ} axial densities of
BECs during the first (a--d) and fourth (e--h) passes through a
semi-permeable defect. The defect is located at $z = 0$, and its
strength was adjusted to keep $\Vd/\mu\sim0.7$.  Oscillation
amplitudes were adjusted to keep $v_0 \sim c_0$.
(a), (e): $a=0.1\,a_0$, $N=1.0\times10^5$, $\mu/h=5\,$Hz,
    $\xi=12.5\,\micron$, $\xi_{s}=16(6)\,\micron$; 
(b), (f): $a=0.5\,a_0$, $N=2.2\times10^5$, $\mu/h=30\,$Hz,
    $\xi=4.9\,\micron$, $\xi_{s}=6.6(2)\,\micron$; 
(c), (g): $a=1.7\,a_0$, $N=2.6\times10^5$, $\mu/h=77\,$Hz,
    $\xi=3.06\,\micron$, $\xi_{s}=2.8(4)\,\micron$; 
(d), (h): $a=5.4\,a_0$, $N=2.2\times10^5$, $\mu/h=144\,$Hz,
    $\xi=2.24\,\micron$, $\xi_{s}=2.5(3)\,\micron$.
The trap frequencies for this data
are $\omega_r = (2\pi)\,240$\,Hz and $\omega_z = (2\pi)\,4.75$\,Hz.
The dashed lines show fits to Eq.~\ref{eqn:soliton}.
We omit the fit in (d) for clarity.
For comparison, the thin dashed line in (a) is only the Gaussian portion of the fit.
Error bars for $\xi_s$ are given by the standard deviation of a collection of images.
} \label{fig:wigglydens}
\end{figure}

We have studied the formation of deep \textit{in situ} density
modulations in BECs for different values of $a$, with the results
shown in Fig.~\ref{fig:wigglydens}. Dipole motion is initiated
after the field is slowly ramped to a desired value near the
scattering length zero-crossing at 544\,G.
Panels (a--e) of Fig.~\ref{fig:wigglydens}
show the cloud after $3/4$ of a complete oscillation.
The defect is switched off after the first pass of the cloud,
and the cloud is imaged after it returns to the center of the trap after another quarter period.
Therefore, $\sim\nosp100$\,ms elapses between the
initial interaction of the cloud with the defect, where the soliton
is created, and imaging. Deep density modulations, consistent with
the formation of stable dark solitons, are observed. For comparison,
panels (e--h) of Fig.~\ref{fig:wigglydens} show the cloud after
passing through the defect 4 times.  The density modulations in this
case appear less monochromatic than in the single pass case,
suggesting the presence of both linear (phonons) and nonlinear
(solitons) excitations. We extract the healing length $\xi_s$ by
fitting the single-pass data in Fig.~\ref{fig:wigglydens} to~\cite{tsuzuki71}
\begin{equation}\label{eqn:soliton}
n (z) = A e^{-z^2/\sigma^2} \left[ 1 - D\,\text{sech}^2 \left( \frac{z-z_0}{\xi_s \sqrt{2}} \right) \right],
\end{equation}
\noindent
where $A$ is the background density, $\sigma$ is the size of the atomic cloud,
$D$ is the depth of the soliton, $z_0$ is the location of the soliton,
and $\xi_s$ is the healing length.  Through a
variational solution of the GPE, we can independently estimate the
healing length $\xi$ using the measured values of $N$, $a$,
$\omega_z$, and $\omega_r$.  The results of this analysis are
reported in Fig.~\ref{fig:wigglydens}.  The average size of the density dips is very nearly the healing length predicted by the GPE estimations, i.e. $\xi_s\sim\xi$.
This observation is consistent with the formation of a downstream dispersive shock consisting of a train of dark solitons as a supersonic BEC crosses a semi-permeable barrier \cite{leboeuf01,pavloff02,Radouani04,Theocharis05,Carretero07,albert08}.

\section{Summary and Future Directions}\label{sec_conc}

We have conducted comprehensive measurements of the dissipation of
superfluid flow in an elongated BEC subject to either a disordered
potential or a single Gaussian defect. By measuring
the velocity and disorder strength-dependent damping parameter, we have
characterized the breakdown of superfluidity of a harmonically
trapped cloud in both the 3D Thomas-Fermi and the
quasi-1D weakly interacting regimes.

Our data largely support the validity of the Landau criterion for a critical velocity above which the superfluid
motion is damped, as long as the criterion is applied \emph{locally}. The local criterion accounts for the
inherent inhomogeneity of trapped gases, as well as density modifications produced by large defects. The only
exception is for attractive defects of relatively large strength, where we find that $v_c$ decreases to
$v_c\sim0.7\,c_0$ for $\Vd/\mu < -0.5$. Dissipation is also found to diminish for velocities greater than $v_+$,
which we associate with reduced excitation of dark solitons and phonons.

Throughout the 3D Thomas-Fermi regime, the damping is found to be well described by a universal relation
depending on the \textit{dimensionless} defect strength $\Vd/\mu$ and velocity $v_0/c_0$. The universal
damping peaks at $v_0/c_0\sim1$ for any $\Vd/\mu$ and scales as $(\Vd/\mu)^2$ for all $\mu$. As $\mu$ decreases,
the peak damping rate decreases as well, consistent with the disappearance of the phonon portion of the
excitation spectrum as $c_0\rightarrow0$. Damping in the quasi-1D regime is qualitatively different. In this
case, we find for fixed \textit{absolute} $\Vd$ and $v_0$ that $\beta$ is independent of $\mu$. In this regime,
damping is accompanied by fragmentation and spreading of the cloud, with the damping monotonically increasing
with $\Vd/E_\mathrm{K}$, where $E_\mathrm{K}$ is the maximum single particle kinetic energy.

\bigskip
A particularly intriguing possibility for the future is to explore
the transport properties of a weakly attractive gas.
{In the case of a disordered potential there exists
the opportunity to study the transport properties of bright
matter-wave solitons \cite{akkermans08} with the prospect of
observing Anderson localization in such systems \cite{kartashov05,
sacha09}.} For a single defect, there is also a possibility for the creation
of coherently split solitons or solitonic Schr\"{o}dinger's cat
states~\cite{streltsov08,weiss09,streltsov09}.

\bigskip
We thank S.~Bhongale, P.~Kakashvili, C.~J.~Bolech, H.~Pu, M.~Albert, P.~Lebeouf, T.~Paul, and
N.~Pavloff for helpful discussions. The NSF, ONR, and the Keck and the Welch Foundations
(C-1133) supported this work.
\bibliography{dipoleosc1}

\end{document}